\documentclass[aps,preprintnumbers,amsmath,amssymb,prl,superscriptaddress,longbibliography,twocolumn]{revtex4-1}
\usepackage{dcolumn}
\usepackage{color}
\usepackage[caption=false]{subfig}
\usepackage{feynmp}
\usepackage{feynmp-auto}
\usepackage{float}
\usepackage{bbm}

\def\comment#1{}

\newcommand{\nc}{\newcommand}
\nc{\beq}{\begin{eqnarray}}
\nc{\eeq}{\end{eqnarray}}
\nc{\scs}{\scriptstyle}
\nc{\setval}{\fmfset{wiggly_len}{3mm} \fmfset{arrow_len}{1.5mm}
	\fmfset{arrow_ang}{13} \fmfset{dash_len}{1.5mm}\fmfpen{0.125mm}
	\fmfset{dot_size}{2thick}}

\usepackage{bm,latexsym,mathrsfs,enumerate,color}
\usepackage[mathcal]{euscript}
\usepackage[breaklinks=true,unicode=true,urlcolor = blue,colorlinks = true,citecolor = blue,linkcolor = blue]{hyperref}
\usepackage{graphicx}
\usepackage{todonotes}
\usepackage{wrapfig}

\renewcommand{\vec}[1]{\bm{#1}}

\def\slashchar#1{\setbox0=\hbox{$#1$}           
	\dimen0=\wd0                                 
	\setbox1=\hbox{/} \dimen1=\wd1               
	\ifdim\dimen0>\dimen1                        
	\rlap{\hbox to \dimen0{\hfil/\hfil}}      
	#1                                        
	\else                                        
	\rlap{\hbox to \dimen1{\hfil$#1$\hfil}}   
	/                                         
	\fi}                                         %

\DeclareMathAlphabet\mathbfcal{OMS}{cmsy}{b}{n}

\def\sigmab{{\mbox{\boldmath $\sigma$}}}
\def\nablab{{\mbox{\boldmath $\nabla$}}}

\newcommand{\angstrom}{\textup{\AA}}

\newcommand{\eus}{EuS-Bi$_2$Se$_3$}
\newcommand{\mbt}{MnBi$_2$Te$_4$}

\usepackage{physics}
\usepackage{xcolor}
\DeclareMathOperator{\sgn}{sgn}

\begin{document}
\title{Finite temperature fluctuation-induced order and responses in magnetic topological insulators}
\author{Marius Scholten}
\affiliation{Institute for Theoretical Solid State Physics, IFW Dresden, Helmholtzstr. 20, 01069 Dresden, Germany}

\author{Jorge I. Facio}
\affiliation{Institute for Theoretical Solid State Physics, IFW Dresden, Helmholtzstr. 20, 01069 Dresden, Germany}

\author{Rajyavardhan Ray}
\affiliation{Institute for Theoretical Solid State Physics, IFW Dresden, Helmholtzstr. 20, 01069 Dresden, Germany}
\affiliation{Dresden Center for Computational Materials Science (DCMS), TU Dresden, 01062 Dresden, Germany}

\author{Ilya M. Eremin}
\affiliation{Institut f\"ur Theoretische Physik III, Ruhr-Universit\"{a}t Bochum, D-44780 Bochum, Germany}

\author{Jeroen van den Brink}
\affiliation{Institute for Theoretical Solid State Physics, IFW Dresden, Helmholtzstr. 20, 01069 Dresden, Germany}
\affiliation{Institute for Theoretical Physics and W\"urzburg-Dresden Cluster of Excellence ct.qmat, TU Dresden, 01069 Dresden, Germany}

\author{Flavio S. Nogueira}
\affiliation{Institute for Theoretical Solid State Physics, IFW Dresden, Helmholtzstr. 20, 01069 Dresden, Germany}
\date{\today }	

\begin{abstract}
We derive an effective field theory model for magnetic topological insulators and predict 
that a magnetic electronic gap persists on the surface for temperatures above the 
ordering temperature of the bulk. Our analysis also applies to interfaces of heterostructures 
consisting of a ferromagnetic and a topological insulator. In order to make quantitative 
predictions for \mbt\, and for EuS-Bi$_2$Se$_3$ heterostructures, we combine the 
effective field theory method with density functional theory and Monte Carlo simulations. 
For \mbt\ we predict an upwards N\'eel temperature shift at the surface up to $15 \%$, 
while the EuS-Bi$_2$Se$_3$ interface exhibits a smaller relative shift. The effective theory 
also predicts induced Dzyaloshinskii-Moriya interactions and 
a topological magnetoelectric effect, both of which feature a finite temperature 
and chemical potential dependence. 
\end{abstract}

\maketitle

\date{\today}

{\it Introduction} --- 
Since the first experimental observation of the quantum anomalous Hall  effect (QAHE) --the appearance of quantized Hall conductance at zero magnetic field-- in thin films of the topological insulator (TI) Bi$_2$Se$_3$ doped with magnetic atoms at temperatures below 1 K \cite{QAHE-review}, magnetic topological materials have been at the scientific forefront both experimentally and theoretically \cite{Yokoyama-Nagaosa-PRB-2010,Garate-Franz-PRL-2010,Tserkovnyak-Loss-PRL-2012,Nogueira-Eremin-PRL-2012-FirstHubbardStratPaper,Nogueira-Eremin-PRB-2014-ChemPotScreening,Nogueira-Eremin-PRB-2013-SemiInsuTransition,Tserkovnyak-Pesin-Loss-PRB-2013-SOMagResponse,Rex-Nogueira-Sudbo-PRB-Jan2016,Li-Katmis-Moodera-PRB-2015,Rex-Nogueira-Sudbo-PRB-July2016}.  
The QAHE requires a three-dimensional TI in which long-range magnetic order breaks the time-reversal symmetry, via ferromagnetic \cite{Nogueira-Katmis-Moodera-Nature-2016-TcShift,Wei-Katmis-Moodera-PRL-2013,Yang-Kapitulnik-PRB-2013-BiSeEuSMagnetoRes,Lee-Katmis-Moodera-Nat-2016} or antiferromagnetic ordering 
\cite{TI-AFMI}.
Driven by the major goal to realize quantization of conductance at room temperature, two distinct directions of material-development have triggered much of the recent experimental progress: the successful fabrication of atomically sharp interfaces of ferromagnetic and topological materials, in particular of EuS and Bi$_2$Se$_3$ \cite{Wei-Katmis-Moodera-PRL-2013,Yang-Kapitulnik-PRB-2013-BiSeEuSMagnetoRes,Nogueira-Katmis-Moodera-Nature-2016-TcShift}, and the growth of intrinsically magnetic TIs like  MnTe(Bi$_2$Te$_3$)$_m$  with $m \geq 1$ \cite{Chulkov-Isaeva,gong2019experimental,Zeugner-Isaeva-ACS-LatticeConst-2019,PhysRevB.100.121104,Otrokov-Chulkov-PRL-MnBiTe-Theory,Isaeva_PhysRevX.9.041065} and MnSb$_2$Te$_4$ \cite{chen2019intrinsic,wimmer2020ferromagnetic} 
as  highly-ordered single-crystals  or their intrinsic heterostructures \cite{Hirahara2017,Hirahara2020}. 

\begin{figure}
	\includegraphics[width=.71\linewidth]{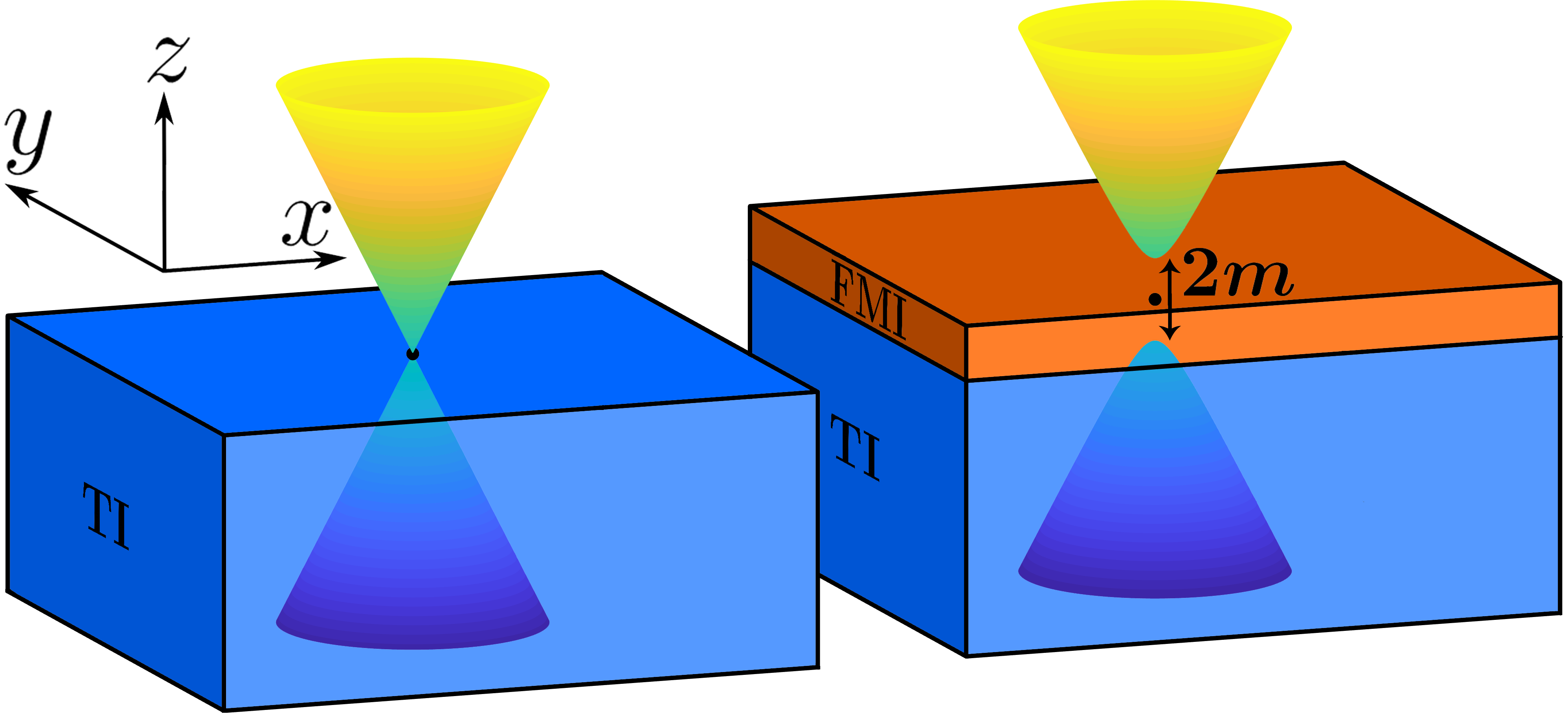}
	\includegraphics[width=.27\linewidth]{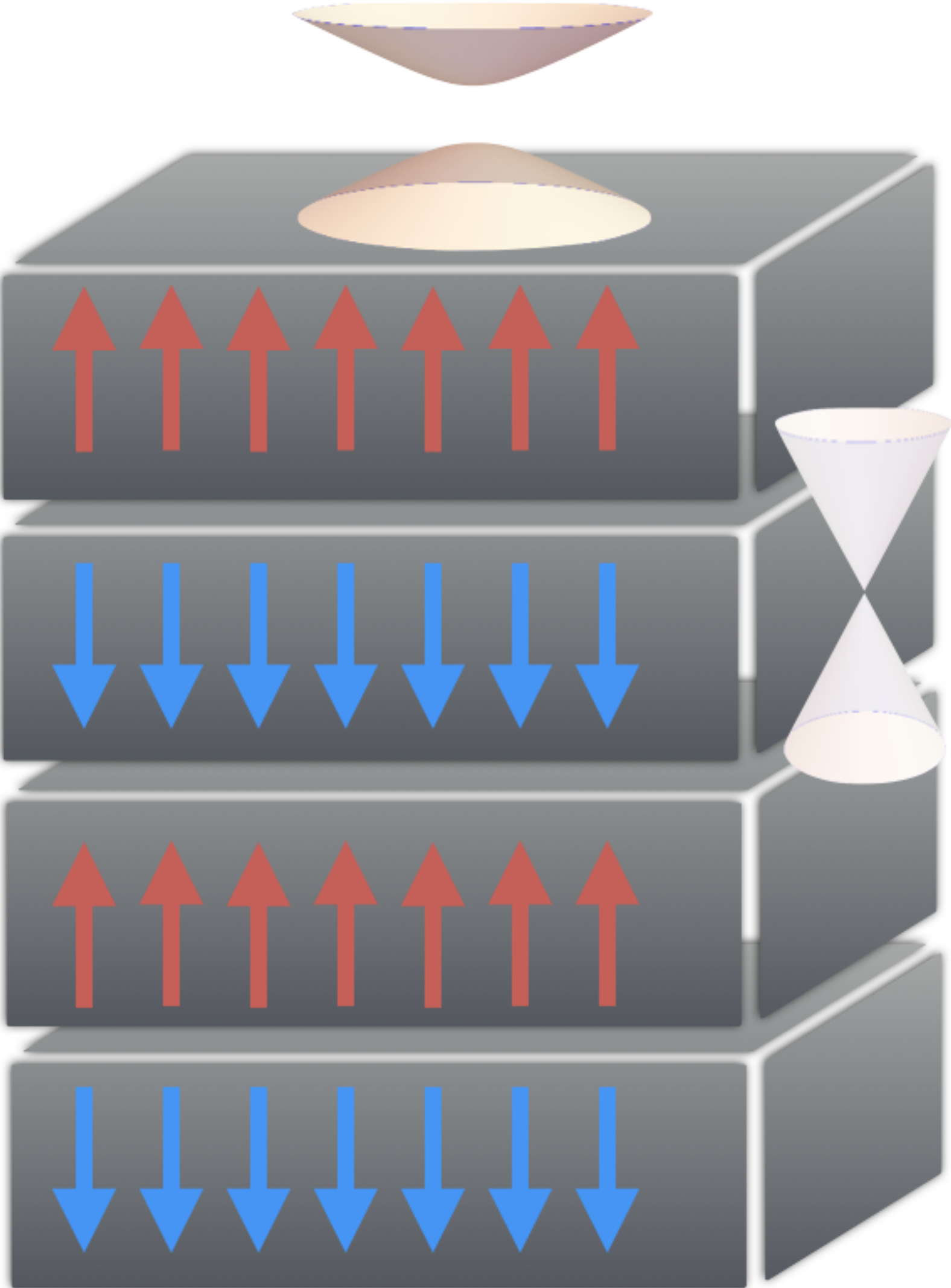}
	\caption{Left: symmetry breaking induced by proximity effect. 
		An exchange coupling is induced across the interface between a FMI and a TI.
		The FMI polarizes the TI surface by proximity-effect and gaps the surface spectrum like, e.g., heterostructures  EuS-Bi$_2$Se$_3$ heterostructues \cite{Wei-Katmis-Moodera-PRL-2013}.
		Right: intrinsic spontaneous symmetry breaking. Here the TI is itself a magnetic insulator like, e.g., MnBi$_2$Te$_4$ \cite{Chulkov-Isaeva}.
	}
\label{TIFMIBilayerfig}
\end{figure}

Conceptually, a perfectly quantized Hall conductivity of $e^2/h$ arises at zero temperature when the magnetization couples to the topological Dirac-type TI surface states, opening up a gap there in which then the chemical potential $\mu$ must lie, see Fig.~\ref{TIFMIBilayerfig}. 
As a bonus, fermionic quantum fluctuations induce a concomitant linear topological magnetoelectric effect (TME), which couples electric fields directly to the magnetization and, {\it  vice versa}, magnetization dynamics to electrical polarization~\cite{Qi-Hughes-Zhang-PRB-2008-TFTTRIInsu,Ryu_PhysRevB.85.045104}.
At zero temperature for $\mu$ outside the gap both Hall conductance and TME fail
to be quantized \cite{Yokoyama-Nagaosa-PRB-2010,Nogueira-Eremin-PRB-2014-ChemPotScreening,Tutschku_PhysRevB.102.205407}, tending to vanish as $\mu$ grows. Nevertheless,  
this particular situation has the interesting feature, that a Dzyaloshinskii Moriya interaction (DMI) between magnetic degrees of freedom emerges, opening the path towards the formation of various skyrmion-like topological magnetic textures at the surface~\cite{Nogueira-Kravchuk-PRB-2018-FMITISkyr,Tiwari2019}, observed also experimentally\cite{Zhang2018}.

The goal to realize the QAHE, TME and possibly even a DMI at room temperature ties into a number of fundamental and practical questions. The first is how in an ideal situation the surface magnetic ordering temperature $T_{\rm c}$ is affected by coupling to the topological edge states -- it has been suggested that this coupling can greatly enhance $T_{\rm c}$ \cite{Nogueira-Katmis-Moodera-Nature-2016-TcShift,Sinova_PhysRevLett.119.027201}. Subsequently the question is how temperature fluctuations affect the conductance quantization, TME and DMI for the different relevant regimes of $\mu$. Particularly interesting would be the existence of a temperature regime in which both DMI and TME are sizeable in which case the TME endows external magnetic and electric fields with novel types of access to DMI-induced skyrmions.

Despite the several theoretical and experimental developments in recent years, 
a number of fundamental questions remain to be answered. For instance, although a remarkable enhancement of $T_{\textrm{c}}$ 
at interfaces of certain ferromagnetic insulators (FMIs) have been reported 
\cite{Nogueira-Katmis-Moodera-Nature-2016-TcShift,Above-room-T-Fe3GeTe2}, recent works 
\cite{Krieger2019,Absence-of-proximity-effect_PhysRevLett.125.226801} question the validity of these 
findings for the specific case of EuS proximate to either Bi$_2$Se$_3$ or (Bi,Sb)$_2$Te$_3$. 
Furthermore, a recent experimental work \cite{Krieger2019} 
indicates that the topological electronic states at the interface do not interact strongly  
with ferromagnetism for the case of a EuS-Bi$_2$Se$_3$ heterostructure.
Additionally, in the family of magnetic TIs MnTe(Bi$_2$Te$_3$)$_m$, different works that do find a surface spectrum gapped below the N\'{e}el temperature also observe the persistence of the gap in the paramagnetic phase  \cite{Chulkov-Isaeva,Isaeva_PhysRevX.9.041065,wu2019natural}, it being unclear whether or not these observations result from the intrinsic magnetism.

Here we develop the finite temperature continuum field theory to address the questions
concerning the magnetic phase transitions and dynamics at finite temperature, and apply these
results to several experimentally relevant materials systems, using Monte-Carlo simulations
(MCS) and density functional theory (DFT) based approaches to obtain quantitative results. Using a
minimal model for the coupling of the Dirac fermions to the magnetic Hamiltonian, we show that in
a temperature window where bulk magnetism is absent, an out-of-plane surface magnetization 
can still be nonzero and induce 
a gap in the Dirac spectrum. As a consequence, the AHE and TME can survive in a certain temperature
range above the bulk $T_{\rm c}$. However, for the experimentally relevant materials and material
combinations (\eus, \mbt),  the coupling to the topological surface states enhance the bulk
$T_{\rm c}$ not more than $15\%$, even under the most favorable conditions. Significant enhancements would 
require using TIs with a much lower Fermi velocity of the Dirac cone. 
In addition, at finite $\mu$ we establish
the existence of a temperature regime displaying both a substantial fluctuation-induced DMI and
TME, even if its Hall conductivity is strongly renormalized, with potentially interesting
consequences for skyrmion manipulation and transport.

{\it Saddle-point and induced order} --- 
To determine the shift in magnetic ordering temperature due to coupling of the magnetic moments to the fluctuating Dirac fermions we consider the 
following minimal model Hamiltonian 
\begin{equation}
\begin{split}
H_{\rm{Dirac}}&=\big[\hbar v_{\textrm{F}}\vec{d}(-i\vb{\grad})-J_0\vec{n}(\vec{r},t)\big]\cdot\pmb{\sigma}-e\phi(\vec{r},t)-\mu,\\
\end{split}
\label{DiracHamGeneral}
\end{equation}
where the Dirac fermions  couple to the magnetization 
via a magnetic exchange interaction $J_0$,  $\pmb{\sigma}$ is the Pauli matrix vector and
$\vec{n}(\vec{r},t)$ the unit vector field representing the magnetization direction at $\vec{r}=(x,y)$. 
The operator $\vec{d}(-i\vb{\grad})$ has the property $\vec{d}^2=-\nabla^2$, where 
$\vb{\grad}=(\partial_x,\partial_y,0)$.  
Additionally, an electric potential $\phi$ has been introduced, 
which includes contributions of an externally applied electric field and an internal 
long-range Coulomb interaction as well. 

The fermionic quantum fluctuations of the 
Dirac Hamiltonian (\ref{DiracHamGeneral}) are accounted for by  the imaginary time path integral, 
\begin{align}
&\mathcal{Z}_{\textrm{F}}=e^{-\beta\mathcal{F}_{\textrm{F}}(\vec{n})}=
\int\mathcal{D}\big[{\Psi}^\dagger,\Psi\big]e^{-\frac{1}{\hbar}S[\Psi^\dagger,\Psi]},\label{PartitionGenDef}\\
&S=\int_0^{\hbar\beta}\dd \tau\int\dd^2r\;{\Psi}^\dagger\big(\hbar\partial_\tau+H_{\rm{Dirac}}\big)\Psi,
\label{PartDef1}
\end{align}
where $\Psi=(\Psi_{\uparrow},\Psi_{\downarrow})^T$ is a spinor of Grassmann fields obtained from 
the second-quantized Hamiltonian \cite{NegeleOrland}. 
The above partition function defines a free energy functional 
$\mathcal{F}_{\textrm{F}}(\vec{n})$ which provides an additional free energy to the one of the magnetic free energy.  As a minimal model leading to the 
latter, we consider the magnetic Hamiltonian, 
\begin{equation}
H_{\textrm{M}}=\int \dd^2r\left[\frac{J}{2}(\nablab\vec{n})^2-\frac{K}{2}n_z^2\right],\label{Eq:MagHamilton} 
\end{equation}
where $J>0$ is the exchange energy and  $K>0$ is the anisotropy energy density (per unit area). The magnetic partition function is given by 
the path integral, 
\begin{equation}
\label{Eq:Z-M}
\mathcal{Z}_{\textrm{M}}=\int\mathcal{D}\vec{n}\mathcal{D}\lambda e^{-\frac{1}{\hbar}S_{\textrm{B}}-\frac{1}{\hbar}\int_{0}^{\hbar\beta}\dd\tau\left[H_{\textrm{M}}+\frac{i}{2}\int \dd^2r\lambda(\vec{n}^2-1)\right]},
\end{equation}
where $S_{\textrm{B}}$ is the Berry phase that arises in the construction of the spin coherent state path integral \cite{Sachdev-book}, and $\lambda$ is 
a Lagrange multiplier field enforcing the constraint $\vec{n}^2=1$. 

Keeping the magnetic fluctuations classical, we obtain the following effective Hamiltonian after integrating 
out the Gaussian fluctuations $n_x$ and $n_y$, along with the fermions, 
\begin{eqnarray}
\label{Eq:Heff-fermions}
H_{\rm eff}&=&k_{\textrm{B}}T\Tr\ln(-J\nabla^2+i\lambda)
\nonumber\\
&-&k_{\textrm{B}}T\Tr\ln[\hbar\partial_\tau-\mu+\hbar v_{\textrm{F}}\vec{d}(-i\vb{\grad})\cdot\sigmab-J_0n_z\sigma_z]
\nonumber\\
&+&\frac{1}{2}\int \dd^2r\left[J(\nablab n_z)^2-Kn_z^2+i\lambda(n_z^2-1)\right]. 
\end{eqnarray}
Variation with respect to $n_z$ leads to the saddle-point equation, 
\begin{equation}
\label{Eq:lambda-fermion}
(\lambda_0-K)n_z=2J_0^2n_zk_{\textrm{B}}T\sum_n\int\frac{\dd^2q}{(2\pi)^2}\frac{1}{(\hbar\omega_n+i\mu)^2+E^2_q},
\end{equation}
where $E_q=\sqrt{(\hbar v_{\textrm{F}}q)^2+m^2}$, $\omega_n=\pi k_{\textrm{B}}T(2n+1)/\hbar$ is a fermionic Matsubara frequency, and we have defined $m^2=J_0^2n_z^2$. Equation 
(\ref{Eq:lambda-fermion}) is solved together with the saddle-point equation for $\lambda$, 
which occurs at $i\lambda=\lambda_{0}$,
\begin{equation}
	\label{Eq:nz}
	n_z^2=1-\frac{2k_\textrm{B}T}{J}\int\frac{\dd^2q}{(2\pi)^2}\frac{1}{q^2+\lambda_{0}/J}. 
\end{equation}
Setting $J_0=0$ in Eq. (\ref{Eq:lambda-fermion}) reduces the saddle-point equations to 
one of a classical ferromagnet with easy-axis anisotropy. In this special case the ordered 
phase immediately implies $\lambda_0=K$ and from Eq. (\ref{Eq:nz}) it is straightforward 
to obtain the critical temperature $T_{\textrm{c}}$ by demanding that $n_z(T_{\textrm{c}})=0$, yielding 
\begin{equation}
	\label{Eq:Tc-No-Fermions}
	k_{\textrm{B}}T_{\textrm{c}}=\frac{\pi J}{\ln(\Lambda_{\textrm{s}}\sqrt{\frac{J}{K}})},
\end{equation}
where a cutoff $\Lambda_{\textrm{s}}\gg \sqrt{K/J}$ has been introduced. 
Note that that the above is consistent with the Mermin-Wagner theorem in the limit 
$K\to 0$. 

Our aim is to calculate the shift of this critical temperature when $J_0\neq 0$, i.e., 
accounting for the fermionic quantum fluctuations. 
After explicitly evaluating the Matsubara sum and integral, Eq. (\ref{Eq:lambda-fermion}) 
becomes
\begin{align}
\lambda_{0}=&-\frac{J_0^2k_{\textrm{B}}T}{2\pi (\hbar v_{\textrm{F}})^2}\left[\ln(1+e^{-\frac{\abs{m}-\mu}{k_{\textrm{B}}T}})+\ln(1+e^{-\frac{\abs{m}+\mu}{k_{\textrm{B}}T}})\right]\nonumber\\
&+K+\frac{J_0^2\Lambda_{\textrm{F}}}{2\pi\hbar v_{\textrm{F}}}\,, \label{Eq:lambda-fermion-1}
\end{align}
where the cutoff $\Lambda_{\textrm{F}}\gg |m|/(\hbar v_{\textrm{F}})$.
The chemical potential is temperature dependent and can be obtained by fixing the particle density.
At the critical temperature $T=\widetilde{T}_{\textrm{c}}$ of the interacting system  
we demand that $m=0$ and obtain 
\begin{equation}
\label{Eq:Lambdac-general}
\lambda_{0\textrm{c}}=K+\frac{J_0^2\Lambda_{\textrm{F}}}{2\pi\hbar v_{\textrm{F}}}-\frac{J_0^2k_{\textrm{B}}\widetilde{T}_{\textrm{c}}}{2\pi (\hbar v_{\textrm{F}})^2}\ln(2+2\cosh(\mu_{\textrm{c}}/k_{\textrm{B}}\widetilde{T}_{\textrm{c}})),
\end{equation}
where $\mu_{\textrm{c}}=\mu(\widetilde{T}_{\textrm{c}})$.
This finally yields the critical temperature shift relative to the situation where  fermions are absent, 
\begin{equation}
\label{Eq:Tc-shift}
\frac{T_{\textrm{c}}-\widetilde{T}_{\textrm{c}}}{\widetilde{T}_{\textrm{c}}} =\frac{k_{\textrm{B}}T_{\textrm{c}}}{2\pi J}\ln\left(\frac{K}{\lambda_{0\textrm{c}}}\right). 
\end{equation}
Since the cutoff is large, it is clear that the argument of the logarithm in Eq. (\ref{Eq:Tc-shift}) is smaller than unity, and therefore 
$\widetilde{T}_{\textrm{c}}>T_{\textrm{c}}$ in all cases. From Eqs. (\ref{Eq:Lambdac-general}) 
and (\ref{Eq:Tc-shift}) we see that smaller values of $v_F$ favor larger shifts of the 
critical temperature.

{\it Material specifics --- }
In order to make quantitative predictions for the material systems of interest, we need to 
determine the values of the coupling parameters and cutoffs appearing in the continuum theory. 
We base such values on {\it ab-initio} and Monte-Carlo calculations, which we find to be consistent with available experimental data. 

Based on DFT for a finite slab, for MnBi$_2$Te$_4$  we find the Fermi velocity as
$\hbar v_{\textrm{F}}= 2.3 \pm 0.2$\,eV\AA\, and the coupling $J_0\approx50\,$meV, whereas for
\eus, we consider $\hbar v_{\textrm{F}} \approx 3.3$\,eV\AA\, and  $J_0\approx54\,$meV
\cite{Wei-Katmis-Moodera-PRL-2013,Eremeev-Chulkov-EuSBiSe-2015,Zhang-Cheng-nature-BiSeFermiVelocity-2009}. Note, the value of 
	$J_0$ was derived from the gap size in the Dirac dispersion assuming its origin is purely magnetic. In reality higher order effects such as hybridization of the surface fermions with the bulk electronic states of MnBi$_2$Te$_4$ or EuS may also influence the gap such that the actual magnetic gap and correspondingly $J_0$ might be smaller.

For the fermionic cutoff $\Lambda_{\textrm{F}}$, we consider that the average surface density of a completely filled band is $1/A$, with $A$ the surface unit cell area.
Since our model describes two surface bands, we fix $\Lambda_{\textrm{F}}$ such that $n(\mu=0)=1/A$. Electron-hole symmetry of the model then implies that $n(\mu\to\infty)=2/A$.

To set the anisotropy $K$ and stiffness $J$ of the magnetization field, we first build 
an anisotropic Heisenberg lattice model which we then map to Eq. (\ref{Eq:MagHamilton}). 
As the
on-site anisotropy depends crucially on the thickness of the FMI layer in the {\eus} system  \cite{Nogueira-Katmis-Moodera-Nature-2016-TcShift}, while the Mn layers in MnBi$_2$Te$_4$ are well separated with relatively small out-of-plane exchange couplings \cite{Chulkov-Isaeva,PhysRevLett.124.197201}, 
we consider the magnetic subsystem to be monolayer thick. The corresponding lattice in both cases is a
two-dimensional triangular lattice spanned by Mn in MnBi$_2$Te$_4$ and Eu on the EuS$(111)$ surface
 \cite{Nogueira-Katmis-Moodera-Nature-2016-TcShift}.
The considered magnetic interactions comprise of nearest neighbor ferromagnetic exchange couplings
$\mathcal{J}$, and an effective on-site out-of-plane anisotropy $\mathcal{K}$. 
$J$ and $K$ follow from $\mathcal{J}$ and $\mathcal{K}$ taking the continuum limit 
\cite{SI}. 

For {\mbt}, we obtain from DFT calculations $\mathcal{K} \approx 0.073$\,meV, and $\mathcal{J}\approx 0.18$\,meV, in good agreement with earlier reported values \cite{Li-McQueeney-PRL-AFMTIMnBi2Te4-ExchangeandAniso-2020}.
For monolayer EuS, on the other hand, $\mathcal{K}$ was obtained by extrapolating the data for the $20$\,nm thick Bi$_2$Se$_3$ layer in the 
{\eus} heterostructures in Ref. [\onlinecite{Nogueira-Katmis-Moodera-Nature-2016-TcShift}] to the
EuS monolayer thickness.
We obtain $\mathcal{K} \approx 0.13$\,meV.
The exchange coupling for the monolayer has been previously estimated to be  $\mathcal{J}= 0.017$ meV \cite{Mauger-Godart-EuS-1986}. It is
interesting that these two material systems cover a broad range of $\mathcal{K}/\mathcal{J}$ (from
$\sim$$0.4$ in the Mn-based compound to $\sim$$7.6$ in the EuS-based system).

Last, the spin cutoff was fixed such the critical temperature of the continuum model (without
fermions) matches the critical temperature of the corresponding Heisenberg model. To obtain the
latter, classical MCS for the lattice model were carried out  (see SI \cite{SI} for
details).
For the {\eus} and {\mbt} system, we obtain $T_{\rm c}^{\rm latt} \approx 5.8\;$K
and $\approx 17.0\;$K, respectively. 
Based on these values, we fix $\Lambda_{\textrm{s}}$ via Eq. (\ref{Eq:Tc-No-Fermions}).

Finally, using Eq. (\ref{Eq:Tc-shift}), we obtain 
 $({T_{\textrm{c}}-\widetilde{T}_{\textrm{c}}})/{\widetilde{T}_{\textrm{c}}} = 10.9\%$ and $14.7\%$ for EuS-Bi$_2$Se$_3$ and  MnBi$_2$Te$_4$, respectively. Note that the compound 
 having smaller Fermi velocity (MnBi$_2$Te$_4$) shows indeed a larger shift of the critical 
 temperature (recall that in both cases $J_0$ is similar).


{\it Fluctuations around the saddle-point} --- 
The question on how the DMI, TME and Hall conductivity evolve with temperature and chemical potential requires considering the effect of magnetic fluctuations in the fermionic  
determinant resulting from integrating out the fermions. 
Introducing
$
\vec{n}(\vec{r},t)=n_z\vu{z}+\delta\vec{n}(\vec{r},t)
$
we can determine the effective action around the saddle-point approximation up to quadratic 
order in the fluctuations. 

\begin{figure}[t]
	\centering
		\includegraphics[width=0.48\columnwidth]{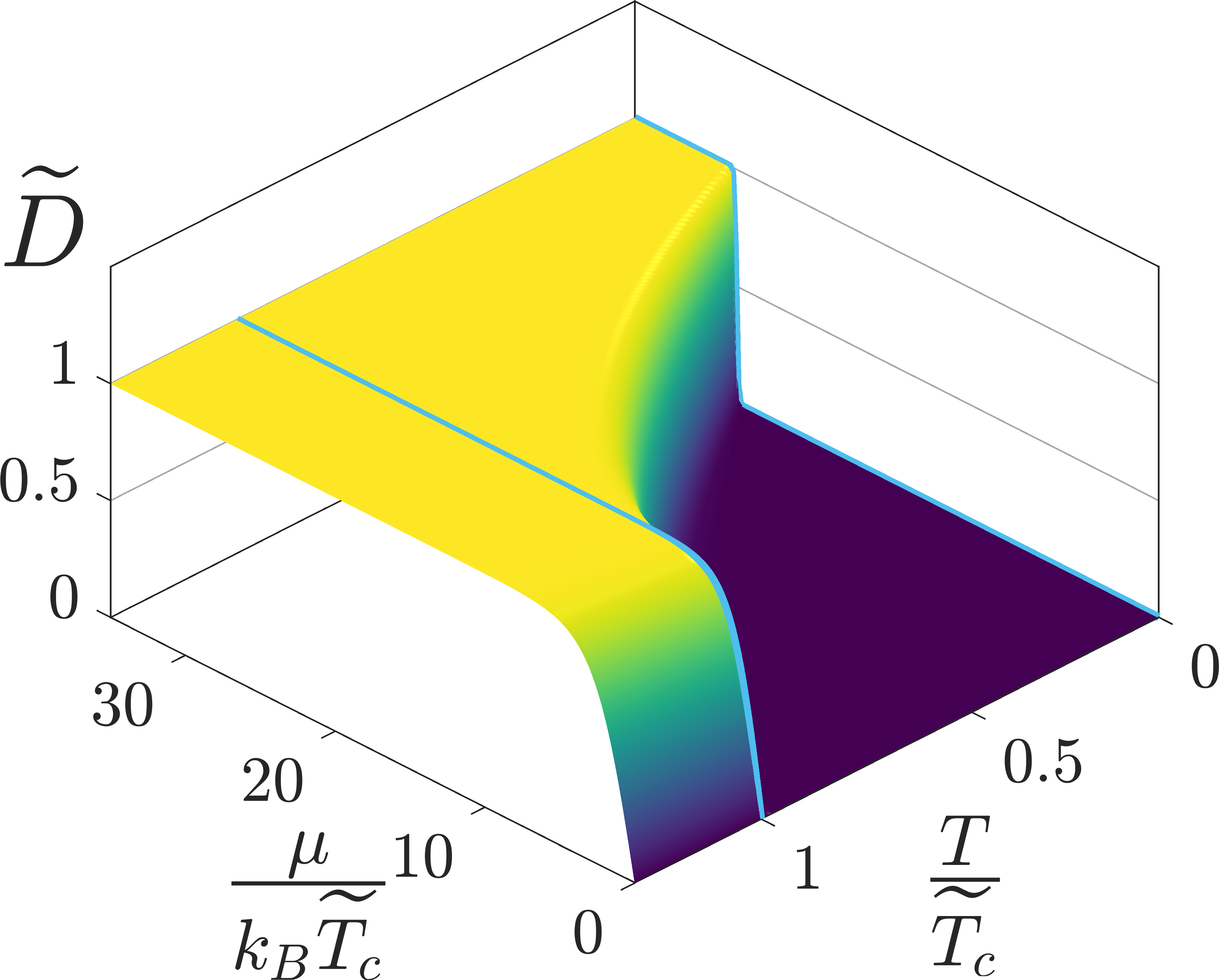}
		\includegraphics[width=0.48\columnwidth]{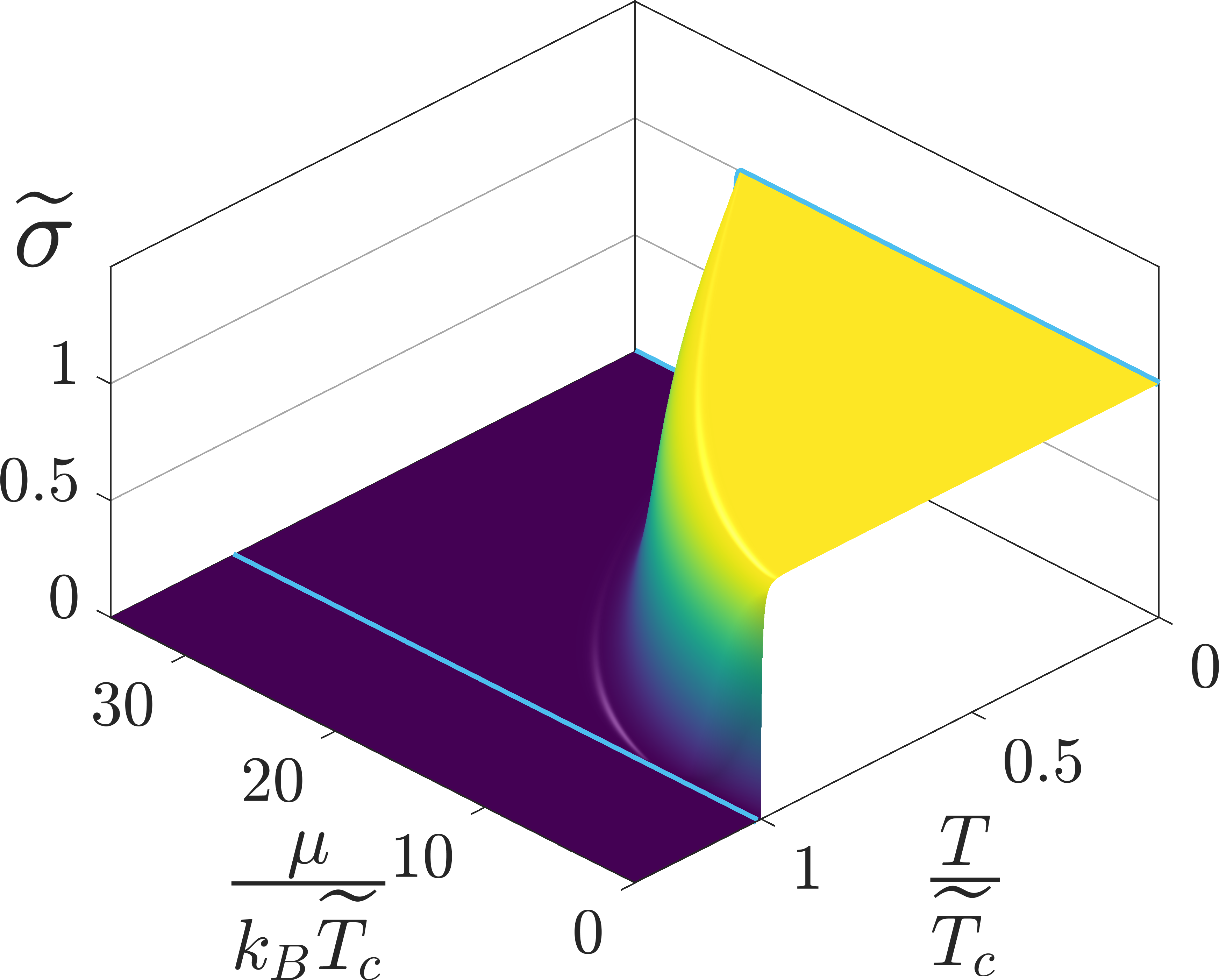}
	\caption{(a) Normalized massive DMI coupling strength $\tilde{D}=D/\bar{D}$ and (b) normalized topological mass $\tilde{\sigma}=\sigma/\bar{\sigma}$, both as a function of temperature and chemical potential. The normalizations are set to $\bar{D}=-J_0^2/4\pi\hbar v_{\textrm{F}}$ and $\bar{\sigma}=e^2/2h$ respectively. 
		The DMI is only present when the chemical potential exceeds the gap (metallic regime). 
		On the contrary the topological mass only exists for a chemical potential inside the gap (insulating regime). 
	}
	\label{FigDMIandCS}
\end{figure}

The reason for the appearance of the DMI term \cite{Nogueira-Kravchuk-PRB-2018-FMITISkyr} is that the magnetization fluctuations effectively 
break the inversion symmetry of our starting Hamiltonian. As a 
consequence, this yields a DMI contribution to the magnetic free energy, 
\begin{equation}
F_{\rm{DMI}}=i\frac{D}{2}\int\dd^2 r\, \delta \vec{n}\cdot\big[\vec{d}(-i\vb{\grad})\cross\delta \vec{n}\big],
\label{NogDMIEn}
\end{equation}
where,
\begin{equation}
D=-\frac{ J_0^2}{4\pi\hbar v_{\textrm{F}}}\frac{\sinh (\beta  \mu )}{\cosh (\beta  \mu )+\cosh (\beta  m)}.
\label{NogDMI}
\end{equation}
It is important to emphasize that the DMI term is not introduced in an {\it ad hoc} way -- it is generated by charge fluctuations coupling to the magnetic moments at the interface.
The DMI vanishes at the neutrality point and is nonzero away from it. This creates the possibility of manipulating the DMI by controlling the chemical potential, for instance by gating.
If we take the mass $m$ to have a mean-field like behavior $m(T)=J_0 \sqrt{1-T/\widetilde{T}_{\textrm{c}}}$ we find the zero temperature value for the DMI
$
D(T\rightarrow 0)=-\frac{J_0^2}{4\pi\hbar  v_{\textrm{F}}}\sgn(\epsilon_{\textrm{F}})H(\abs{\epsilon_{\textrm{F}}}-\abs{J_0})
$
which demonstrates that the DMI kicks in when the Fermi energy $\epsilon_{\rm{F}}$ surpasses a threshold given by the exchange coupling constant $J_0$. This feature of the generated DMI can also be seen in Fig. \ref{FigDMIandCS} (a) where we show the whole temperature range for different values of the chemical potential with an estimated zero temperature exchange coupling constant of $J_0/k_{\textrm{B}}\widetilde{T}_{\textrm{c}}\approx29.1$ based on our findings for the {\mbt} system. We further see that the lower the temperature gets, the narrower the range for the chemical potential becomes in which there still is a finite DMI. Moreover, we see that the step-function behavior of the generated DMI at zero temperature approximately extends to the whole temperature range as the DMI is only present when the chemical potential exceeds the magnetic gap, meaning it exists only in the metallic regime.

Finally, we determine 
the fluctuation-induced effective Chern-Simons (CS) action, 
\begin{equation}
S_{\rm{eff}}^{\rm{cs}}=\frac{\sigma}{2}\int\dd (ct)\int\dd^2 r\;\epsilon^{\mu\nu\lambda}\mathcal{A}_\mu\partial_\nu\mathcal{A}_\lambda,
\label{MyChernSimonsAction}
\end{equation}
where we have defined the covariant three-potential 
$\mathcal{A}_\mu=\left(\frac{\phi}{c},\pm\frac{J_0}{ev_{\textrm{F}}}\vec{d}(\delta\vec{n})\right)$. 
The electric potential enters in the time component as usual and the magnetization fluctuations $\delta\vec{n}$ act as the vector potential $\vb{A}$ in the spatial components. Note that the "$\pm$" applies to the different choices for the vector $\vec{d}$.  The coefficient $\sigma$ arising in Eq. (\ref{MyChernSimonsAction}) leads to the gap in magnetic susceptibility, in a mechanism closely related to the well known topologically massive photons in a Maxwell-Chern-Simons theory \cite{Deser-Jackiw-Templeton-PRL-1982-3DGauge}. In our case this topological mass is given by,
\begin{gather}
	\sigma=\frac{e^2}{2h}\frac{\sinh (\beta  m)}{\cosh (\beta  \mu )+\cosh (\beta  m)}.
\end{gather}
We recall here that the topological mass arising in the effective free energy is in 
general not identical to the Hall conductivity --- these quantities differ, for instance, 
in the metallic regime \cite{Nogueira-Eremin-PRB-2014-ChemPotScreening,Tutschku_PhysRevB.102.205407}, something that is more easily seen in the zero temperature limit. Indeed, for $T=0$ the topological mass and Hall conductivity are given by \cite{Nogueira-Eremin-PRB-2014-ChemPotScreening} $\sigma(T=0)=e^2{\rm sgn}(J_0)H(|J_0|-\epsilon_{\rm{F}})/(2h)$ and $\sigma_{xy}(T=0)=e^2\{[{\rm sgn}(J_0)-J_0/\epsilon_{\rm{F}}]H(|J_0|-\epsilon_{\textrm{F}})+J_0/\epsilon_{\rm{F}}\}/(2h)$, respectively. These zero temperature expressions involving the Heaviside step function are identical only when $\epsilon_{\rm{F}}<|J_0|$. In fact, $\sigma(T=0)$ vanishes in the metallic regime while $\sigma_{xy}$ is nonzero. This occurs because the Hall conductivity is calculated from the Kubo formula where one first takes the limit $\vec{q}\to 0$ and then $\omega\to 0$, while in case of the topological mass these limits are taken simultaneously. The CS action 
(\ref{MyChernSimonsAction}) contains the TME contribution to the free energy, 
\begin{equation}
	F_{\rm{TME}}=i\frac{J_0\sigma}{ev_{\textrm{F}}}\int\dd^2 r\, \delta \vec{n}\cdot\big[\vu{z}\cross\vec{d}(-i\vb{\grad})\phi\big],
	\label{TMEterm}
\end{equation}

\begin{figure}[t]
	\includegraphics[width=.98\columnwidth]{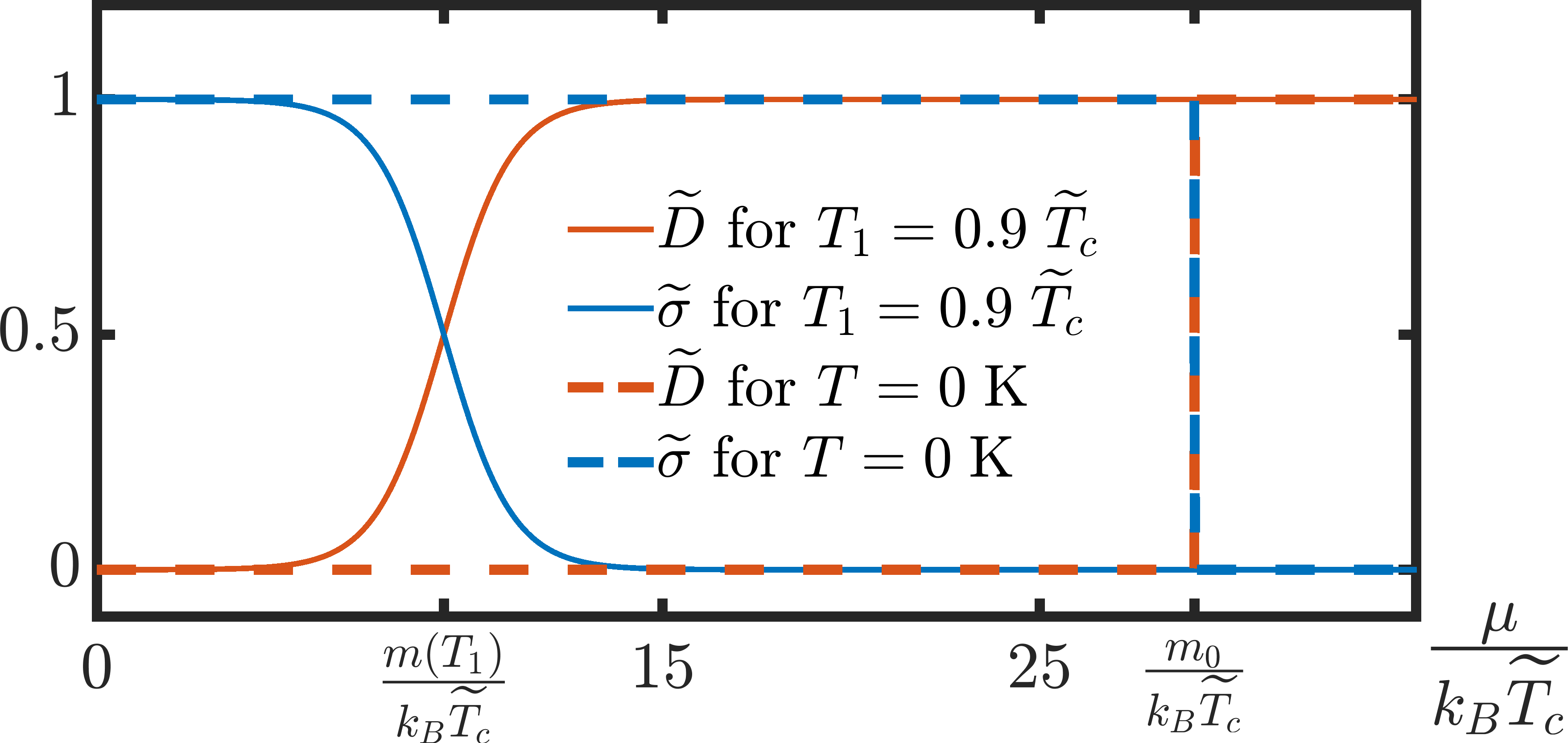}
	\caption{Crossover of the normalized topological mass $\tilde{\sigma}$ and DMI coupling strength $\tilde{D}$ as a function of the chemical potential at different temperatures.  
		For a vanishing temperature both quantities are step functions. 
		Increasing the temperature causes the step functions to smear out,  
		making coexistence possible. The two quantities  always sum up to unity at a given value for the chemical potential.}
	\label{Crossoverfig}
\end{figure}

To compare its features to the ones of the DMI we also illustrate its dependency on temperature and
chemical potential in Fig. \ref{FigDMIandCS} (b). Once more it shows that at zero temperature we
have a step function behavior which also approximately extends to finite temperatures. Consequently,
the topological mass only exists when the chemical potential lies inside the magnetic gap and is
nearly quantized in the bordering regions resulting into plateaus. In comparison to the generated
DMI we can see in Fig. \ref{Crossoverfig} that there is a very narrow region where both functions
overlap. As a result, the desired simultaneous occurrence of the DMI and the CS action requires fine tuning.

However, upon closer inspection, a different connection between the two terms appears. It turns out that the temperature functions inside both terms complement each other almost perfectly in a temperature and chemical potential plot, as also can be seen for exemplary temperatures in Fig. \ref{Crossoverfig}. The two functions are adding up to one creating a plateau that even traverses the chasm that both functions showed individually in the vicinity where the chemical potential crossed the magnetic gap. This means that at the time one of the terms diminishes, the respective other term grows in size equal to the loss of the other, creating a direct correspondence between them.
We point out that besides the DMI and CS terms, other interesting terms appear in the effective action, 
as is shown in explicitly in the SI \cite{SI}. 

{\it Conclusion} --- We have considered a minimal model for magnetic topological insulators that 
capture a wealth of interesting properties of the surface of materials like 
MnTe(Bi$_2$Te$_3$)$_m$  with $m \geq 1$  \cite{Chulkov-Isaeva,Zeugner-Isaeva-ACS-LatticeConst-2019,Otrokov-Chulkov-PRL-MnBiTe-Theory,Isaeva_PhysRevX.9.041065} and MnSb$_2$Te$_4$ \cite{chen2019intrinsic,wimmer2020ferromagnetic}, and also 
at interfaces of heterostructures EuS-Bi$_2$Se$_3$ \cite{Wei-Katmis-Moodera-PRL-2013,Yang-Kapitulnik-PRB-2013-BiSeEuSMagnetoRes,Nogueira-Katmis-Moodera-Nature-2016-TcShift}. An important main result of our analysis is the prediction of the survival of 
the electronic gap on the surface/interface for temperatures above the bulk ordering 
temperature. In order to provide quantitative results to be compared with experiment, 
we have combined the effective field theory analysis with DFT and MCS results 
applied specifically to MnBi$_2$Te$_4$ and the bilayer system EuS-Bi$_2$Se$_3$. 
We also predict that temperature dependent DMI and TME terms are induced by fluctuations. 
The latter may give rise to new magnetic phenomena at 
the surface of magnetic TIs, including the interesting possibility of 
manipulating skyrmions by external or internal electrical fields.  

\begin{acknowledgments}
	JvdB acknowledges support by the DFG through the W\"urzburg-Dresden Cluster of Excellence on Complexity and Topology in Quantum Matter – ct.qmat (EXC 2147, project-id 39085490) and through SFB 1143 (project-id 247310070). JIF would like to 
	thank the support of the Alexander von Humboldt Foundation. 
	I.E. would like to thank the DFG Priority Program SPP 1666, “Topological Insulators,” under Grant No.ER 463/9. The authors thank Ulrike Nitzsche for technical assistance.
\end{acknowledgments}

\bibliography{literature}

\clearpage
\newpage
\onecolumngrid

\appendix

\section{\large Supplemental Materials}
\label{SuppInfo-AppChap}

\section{Saddle-point and induced order}
\label{SuppInfo-SP-SubSec}
The starting point of our field theory is the partition function from the main text, consisting of two parts, namely the magnetic and fermionic contributions. We calculate this partition function in the imaginary time path integral formalism integrating out the fermionic and bosonic fields respectively. 
Keeping the magnetic fluctuations classical, we look at the magnetic partition function and define a corresponding magnetic action
\begin{equation}
\label{Eq:App:Z-M}
\mathcal{Z}_{\textrm{M}}=\int\mathcal{D}\vec{n}\mathcal{D}\lambda e^{-\beta \left[H_{\textrm{M}}+\frac{i}{2}\int d^2r\lambda(\vec{n}^2-1)\right]}=\int\mathcal{D}\vec{n}\mathcal{D}\lambda e^{-\hat{S}_{\textrm{M}}},
\end{equation}
where the magnetic action is given by
\begin{equation}
\begin{split}
\label{Eq:App:S-M}
\hat{S}_{\textrm{M}}=&\frac{\beta}{2}\int\dd^2r\left[J(\nablab n_z)^2-K n_z^2+i\lambda(n_z^2-1)\right]\\
&+\frac{\beta}{2}\int\dd^2r\left[n_a(-J\nabla^2+i\lambda)n_a\right].
\end{split}
\end{equation}
Here we already split up the different components of the magnetization into in-plane and out-of-plane components, where the sum over $a$ denotes the $x$ and $y$ components, such that in the next step we can perform the path integration over the Gaussian fluctuations $n_x$ and $n_y$ yielding the effective action
\begin{equation}
\begin{split}
\label{App:Eq:S-M,Eff}
\hat{S}_{\textrm{M},\textrm{eff}}=&\frac{\beta}{2}\int\dd^2r\left[J(\nablab n_z)^2-K n_z^2+i\lambda(n_z^2-1)\right]\\
&+\Tr\ln(-J\nabla^2+i\lambda).\\
\end{split}
\end{equation}
Doing the same for the fermionic partition function, just by integration over the fermionic fields we get
\begin{eqnarray}
\label{App:Eq:Seff-fermions}
\hat{S}_{\rm eff}&=&-\Tr\ln[\hbar\partial_\tau-\mu+\hbar v_{\textrm{F}}\vb{d}(-i\vb{\grad})\cdot\sigmab-J_0n_z\sigma_z]. 
\end{eqnarray}
To note is the additional minus sign in front, resulting from the integration over fermionic fields instead of bosonic ones. Furthermore, the two so defined functional traces differ from each other, which will soon become apparent.
Combining the two individual contributions we get the total effective action
\begin{eqnarray}
\label{App:Eq:Seff-Mag+fermions}
\hat{S}_{\rm eff}&=&\Tr\ln(-J\nabla^2+i\lambda)
\nonumber\\
&-&\Tr\ln[\hbar\partial_\tau-\mu+\hbar v_{\textrm{F}}\vb{d}(-i\vb{\grad})\cdot\sigmab-J_0n_z\sigma_z]
\nonumber\\
&+&\frac{\beta}{2}\int d^2r\left[J(\nablab n_z)^2-Kn_z^2+i\lambda(n_z^2-1)\right],
\end{eqnarray}
which implies the effective Hamiltonian given in the main text.\\
In the next step we apply the saddle-point approximation to the remaining fields to be integrated over
\begin{equation}
\begin{split}
\label{App:Eq:SPApprox}
n_z(\vb{r})&=\tilde{n}_z+\delta n_z(\vb{r}),\\
\lambda(\vb{r})&=\tilde{\lambda}+\delta \lambda(\vb{r}),\\
\end{split}
\end{equation}
leading to the saddle point effective action
\begin{eqnarray}
\label{App:Eq:Seff,sp-Mag+fermions}
\hat{S}_{\rm eff, sp}&=&\Tr\ln(-J\nabla^2+i\tilde{\lambda})
\nonumber\\
&-&\Tr\ln[\hbar\partial_\tau-\mu+\hbar v_{\textrm{F}}\vb{d}(-i\vb{\grad})\cdot\sigmab-J_0\tilde{n}_z\sigma_z]
\nonumber\\
&+&\frac{\beta V}{2}\left[i\tilde{\lambda}(\tilde{n}_z^2-1)-K \tilde{n}_z^2\right].
\end{eqnarray}
To keep the notation short we redefine $\tilde{\lambda}=\lambda$ and $\tilde{n}_z=n_z$ by dropping the tilde over both quantities. With this we calculate the functional traces
\begin{equation}
\begin{split}
\label{App:Eq1:S-M,TrLn}
\Tr\ln(-J\nabla^2+i\lambda)=\int\dd^2r \mel**{\vb{r}}{\ln(-J\nabla^2+i\lambda)}{\vb{r}},
\end{split}
\end{equation}
where
\begin{equation}
\begin{split}
\label{App:Eq2:S-M,TrLn}
\mel**{\vb{r}'}{\ln(-J\nabla^2+i\lambda)}{\vb{r}}&=\int\frac{\dd^2p}{(2\pi)^2}\int\frac{\dd^2q}{(2\pi)^2}\braket{\vb{r}'}{\vb{p}}\mel**{\vb{p}}{\ln(-J\nabla^2+i\lambda)}{\vb{q}}\braket{\vb{q}}{\vb{r}}\\
&=\int\frac{\dd^2p}{(2\pi)^2}\int\frac{\dd^2q}{(2\pi)^2}e^{i\vb{r}'\cdot\vb{p}}\mel**{\vb{p}}{\ln(-J\nabla^2+i\lambda)}{\vb{q}}e^{-i\vb{r}\cdot\vb{q}}\\
&=\int\frac{\dd^2q}{(2\pi)^2}e^{i\vb{q}\cdot(\vb{r}'-\vb{r})}\ln(Jq^2+i\lambda),
\end{split}
\end{equation}
which upon reinsertion gives
\begin{equation}
\begin{split}
\label{App:Eq3:S-M,TrLn}
\Tr\ln(-J\nabla^2+i\lambda)=V\int\frac{\dd^2q}{(2\pi)^2}\ln(Jq^2+i\lambda).
\end{split}
\end{equation}
Similarly, the functional trace from the integration over the fermionic fields is given by
\begin{equation}
\begin{split}
\label{App:Eq1:Seff-Tr-fermions}
\Tr\ln[\hbar\partial_\tau-\mu+\hbar v_{\textrm{F}}\vb{d}(-i\vb{\grad})\cdot\sigmab-J_0n_z\sigma_z]=\sum_{\sigma}\int_0^{\hbar\beta}\dd\tau \int\dd^2r \mel**{\vb{r},\tau,\sigma}{\ln[\hbar\partial_\tau-\mu+\hbar v_{\textrm{F}}\vb{d}(-i\vb{\grad})\cdot\sigmab-J_0n_z\sigma_z]}{\vb{r},\tau,\sigma},
\end{split}
\end{equation}
where
\begin{equation}
\begin{split}
\label{App:Eq2:Seff-Tr-fermions}
&\mel**{\vb{r}',\tau',\sigma'}{\ln[\hbar\partial_\tau-\mu+\hbar v_{\textrm{F}}\vb{d}(-i\vb{\grad})\cdot\sigmab-J_0n_z\sigma_z]}{\vb{r},\tau,\sigma}\\
=&\frac{1}{\hbar^2\beta^2}\sum_{n,m}\int\frac{\dd^2p}{(2\pi)^2}\int\frac{\dd^2q}{(2\pi)^2}\bra{\sigma'}\braket{\vb{r}',\tau'}{\vb{p},m}\mel**{\vb{p},m}{\ln[\hbar\partial_\tau-\mu+\hbar v_{\textrm{F}}\vb{d}(-i\vb{\grad})\cdot\sigmab-J_0n_z\sigma_z]}{\vb{q},n}\braket{\vb{q},n}{\vb{r},\tau}\ket{\sigma}\\
=&\frac{1}{\hbar\beta}\sum_n\int\frac{\dd^2q}{(2\pi)^2}e^{i\vb{q}\cdot(\vb{r}'-\vb{r})-i\omega_n(\tau'-\tau)}\mel**{\sigma'}{\ln[i\hbar \omega_n-\mu+\hbar v_{\textrm{F}}\vb{d}(\vb{q})\cdot\sigmab-J_0n_z\sigma_z]}{\sigma},
\end{split}
\end{equation}
with the fermionic Matsubara frequencies $\omega_n= (2n+1)\pi/\hbar\beta$. Therefore the functional trace yields
\begin{equation}
\begin{split}
\label{App:Eq3:Seff-Tr-fermions}
\Tr\ln[\hbar\partial_\tau-\mu+\hbar v_{\textrm{F}}\vb{d}(-i\vb{\grad})\cdot\sigmab-J_0n_z\sigma_z]&=V\int\frac{\dd^2q}{(2\pi)^2}\sum_{n,\sigma} \mel**{\sigma}{\ln[i\hbar \omega_n-\mu+\hbar v_{\textrm{F}}\vb{d}(\vb{q})\cdot\sigmab-J_0n_z\sigma_z]}{\sigma}\\
&=V\int\frac{\dd^2q}{(2\pi)^2}\sum_{n,\sigma}\ln(\sigma E_q-\mu+i\hbar\omega_n),
\end{split}
\end{equation}
where the sum runs over $\sigma\in \{-1,1\}$ and we defined $E_q=\sqrt{\hbar^2 v_{\textrm{F}}^2q^2+m^2}$ with the mass $m=J_0n_z$.
Using this we can now formulate the saddle point equations by variation of the effective action with respect to $n_z$ 
\begin{equation}
\label{App:Eq:lambda-fermion}
(\lambda_0-K)n_z=2J_0^2n_zk_{\textrm{B}}T\sum_n\int\frac{d^2q}{(2\pi)^2}\frac{1}{(\hbar\omega_n+i\mu)^2+E^2_q},
\end{equation}
and with respect to $\lambda_{0}=i\lambda$
\begin{equation}
\label{App:Eq:nz}
n_z^2=1-\frac{2k_\textrm{B}T}{J}\int\frac{d^2q}{(2\pi)^2}\frac{1}{q^2+\lambda_{0}/J}. 
\end{equation}
Setting $J_0=0$ in Eq. (\ref{App:Eq:lambda-fermion}) reduces the saddle-point equations to one of a classical ferromagnet with easy-axis anisotropy. In this special case the ordered phase immediately implies $\lambda_0=K$ and from Eq. (\ref{App:Eq:nz}) it is straightforward to obtain the critical temperature $T_{\textrm{c}}$ by demanding that $n_z(T_{\textrm{c}})=0$, yielding 
\begin{equation}
\label{App:Eq:Tc-No-Fermions}
k_{\textrm{B}}T_{\textrm{c}}=\frac{2\pi J}{\ln(1+\frac{J\Lambda_{\textrm{s}}^2}{K})}\approx\frac{\pi J}{\ln(\Lambda_{\textrm{s}}\sqrt{\frac{J}{K}})},
\end{equation}
where a cutoff $\Lambda_{\textrm{s}}\gg \sqrt{K/J}$ has been introduced.
Our aim is to calculate the shift of this critical temperature when $J_0\neq 0$, i.e. 
accounting to the fermionic quantum fluctuations. 
After explicitly evaluating the Matsubara sum and integral, Eq. (\ref{App:Eq:lambda-fermion}) 
becomes,
\begin{align}
	\lambda_{0}=&K+\frac{J_0^2\Lambda_{\textrm{F}}}{2\pi\hbar v_{\textrm{F}}}-\frac{J_0^2k_{\textrm{B}}T}{2\pi (\hbar v_{\textrm{F}})^2}\left[\ln(1+e^{-\frac{\abs{m}-\mu}{k_{\textrm{B}}T}})+\ln(1+e^{-\frac{\abs{m}+\mu}{k_{\textrm{B}}T}})\right],\label{App:Eq:lambda-fermion-1}
\end{align}
where we have assumed that the cutoff $\Lambda_{\textrm{F}}\gg |m|/(\hbar v_{\textrm{F}})$.
The chemical potential is temperature dependent and can be obtained by fixing the particle density.
At the critical temperature $T=\widetilde{T}_{\textrm{c}}$,  we demand that $m=0$ and obtain 
\begin{equation}
\label{App:Eq:Lambdac-general}
\lambda_{0\textrm{c}}=K+\frac{J_0^2\Lambda_{\textrm{F}}}{2\pi\hbar v_{\textrm{F}}}-\frac{J_0^2k_{\textrm{B}}\widetilde{T}_{\textrm{c}}}{2\pi (\hbar v_{\textrm{F}})^2}\ln(2+2\cosh(\mu_{\textrm{c}}/k_{\textrm{B}}\widetilde{T}_{\textrm{c}})),
\end{equation}
where $\mu_{\textrm{c}}=\mu(\widetilde{T}_{\textrm{c}})$.
Analogously to Eq. (\ref{App:Eq:Tc-No-Fermions}) we then get
\begin{equation}
\begin{split}
\label{App:Eq:TildeT_c}
k_{\textrm{B}}\widetilde{T}_{\textrm{c}}=\frac{2\pi J}{\ln(1+\frac{J\Lambda_{\textrm{s}}^2}{\lambda_{0c}})}.
\end{split}
\end{equation}
This finally yields the critical temperature shift relative to the situation where  fermions are absent, 
\begin{equation}
\label{App:Eq:Tc-shift}
\frac{T_{\textrm{c}}-\widetilde{T}_{\textrm{c}}}{\widetilde{T}_{\textrm{c}}} =\frac{k_{\textrm{B}}T_{\textrm{c}}}{2\pi J}\ln(\frac{1+\frac{J\Lambda_{\textrm{s}}^2}{\lambda_{0c}}}{1+\frac{J\Lambda_{\textrm{s}}^2}{K}})\approx \frac{k_{\textrm{B}}T_{\textrm{c}}}{2\pi J}\ln\left(\frac{K}{\lambda_{0c}}\right). 
\end{equation}
Since the cutoff is large, it is clear that the argument of the logarithm in Eq. (\ref{App:Eq:Tc-shift}) is smaller than unity, and therefore $\widetilde{T}_{\textrm{c}}>T_{\textrm{c}}$.

\section{Surface particle density and fermionic cutoff}
The fermionic cutoff $\Lambda_{\textrm{F}}$ can be determined from the two dimensional surface particle density given by
\begin{equation}
\begin{split}
\label{App:Eq1:nDen}
n=\int\frac{\dd^2q}{(2\pi)^2}\sum_Ef(E),
\end{split}
\end{equation}
where in our case the energies are $E=\pm E_q-\mu$ resulting from the Dirac Hamiltonian. By insertion of these energies into Eq. (\ref{App:Eq1:nDen}) the surface particle density becomes
\begin{equation}
\begin{split}
\label{App:Eq2:nDen}
n&=\int\frac{\dd^2q}{(2\pi)^2}\left(1+f(E_q-\mu)-f(E_q+\mu)\right)\\
&=\frac{\Lambda_{\textrm{F}}^2}{4\pi}+I,
\end{split}
\end{equation}
where $I$ is given by
\begin{equation}
\begin{split}
\label{App:Eq3:nDen}
I=\frac{\text{Li}_2\left[-e^{-\beta(\mu+\abs{m})}\right]-\text{Li}_2\left[-e^{\beta(\mu-\abs{m})}\right]}{2\pi\hbar^2v_{\textrm{F}}^2\beta^2}+\frac{\abs{m}}{2\pi\hbar^2v_{\textrm{F}}^2\beta}\left[\ln(1+e^{\beta(\mu-\abs{m})})-\ln(1+e^{-\beta(\mu+\abs{m})})\right],
\end{split}
\end{equation}
with the notation $\text{Li}_n\left[x\right]$ for the polylogarithm.

We now consider that the average surface density of a completely filled band is $1/A$, with the surface unit cell area  $A$. Since our model describes two surface bands, we fix $\Lambda_{\textrm{F}}$ such that $n(\mu=0)=1/A$. At $\mu=0$ the Integral $I$ vanishes, giving an expression for the fermionic cutoff $\Lambda_{\textrm{F}}=\sqrt{4\pi n(\mu=0)}$.

\section{Mapping between lattice and continuum spin models}
We start with an anisotropic Heisenberg model on a two-dimensional triangular lattice:
\begin{equation}
\begin{split}
H=-\mathcal{J}\sum_{<i,j>}\vb{S}_i\cdot\vb{S}_j-\mathcal{K}\sum_iS_{i,z}^2,
\label{AppHamHeisenberg}
\end{split}
\end{equation}
where, $\mathcal{J}\geq 0$ is the nearest neighbor ferromagnetic Heisenberg exchange coupling and $\mathcal{K}$ is the
on-site magnetic anisotropy.
Introducing $\vb{n}(\vb{r}_i)=\vb{S}_i/S$ and $\Delta\vb{R}=\vb{r}_j-\vb{r}_i$ as the distance vector between lattice site $i$ and $j$ the Hamiltonian becomes
\begin{equation}
\begin{split}
H=-\frac{\mathcal{J}S^2}{2}\sum_{i,\Delta\vb{R}}\vb{n}(\vb{r}_i)\cdot\vb{n}(\vb{r}_i+\Delta\vb{R})-\mathcal{K}S^2\sum_in_{z}^2(\vb{r}_i).
\label{AppHamHeisenberg2}
\end{split}
\end{equation}
The (isotropic) Heisenberg exchange term is given by the first part: 
\begin{equation}
\begin{split}
H_{\rm  ex}=-\frac{\mathcal{J}S^2}{2}\sum_{i,\Delta\vb{R}}\vb{n}(\vb{r}_i)\cdot\vb{n}(\vb{r}_i+\Delta\vb{R}).
\label{AppHamJ}
\end{split}
\end{equation}
Here, the scalar product can be written as
\begin{equation}
\begin{split}
\vb{n}(\vb{r}_i)\cdot\vb{n}(\vb{r}_i+\Delta\vb{R})&=1-\frac{1}{2}\left[1-2\vb{n}(\vb{r}_i)\cdot\vb{n}(\vb{r}_i+\Delta\vb{R})+1\right]\\
&=1-\frac{1}{2}\left[\vb{n}(\vb{r}_i)^2-2\vb{n}(\vb{r}_i)\cdot\vb{n}(\vb{r}_i+\Delta\vb{R})+\vb{n}(\vb{r}_i+\Delta\vb{R})^2\right]\\
&=1-\frac{1}{2}\left[\vb{n}(\vb{r}_i)-\vb{n}(\vb{r}_i+\Delta\vb{R})\right]^2\\
&\approx 1-\frac{1}{2}\left[(\Delta\vb{R}\cdot\vb{\grad})\vb{n}(\vb{r}_i)\right]^2,
\label{Appcdot}
\end{split}
\end{equation}
where, in the last step, we retain terms up to $\order{1}$ in $\Delta\vb{R}$.
Therefore, the exchange term becomes 
\begin{equation}
\begin{split}
H_{\rm  ex} &=-\frac{\mathcal{J}S^2}{2}\sum_{i,\Delta\vb{R}}\left(1-\frac{1}{2}\left[(\Delta\vb{R}\cdot\vb{\grad})\vb{n}(\vb{r}_i)\right]^2\right)\\
&=\frac{\mathcal{J}S^2}{4}\sum_{i,\Delta\vb{R}}\left[(\Delta\vb{R}\cdot\vb{\grad})\vb{n}(\vb{r}_i)\right]^2+\textrm{const.}\\
&=\frac{\mathcal{J}S^2}{4}\sum_{i,j,\Delta\vb{R}}\left[(\Delta\vb{R}\cdot\vb{\grad})n_j(\vb{r}_i)\right]^2+\textrm{const.}
\label{AppHamJ2}
\end{split}
\end{equation}
After carrying out the sum over $\Delta\vb{R}$ for our triangular lattice, we have 
\begin{equation}
\begin{split}
H_{\rm  ex} &=\frac{\mathcal{J}S^2}{4}\sum_{i,j,\Delta\vb{R}}\left[(\Delta\vb{R}\cdot\vb{\grad})n_j(\vb{r}_i)\right]^2\\
&=\frac{\mathcal{J}S^2}{4}\sum_{i,j}3a^2\left(\vb{\grad}n_j(\vb{r}_i)\right)^2\\
&=\frac{3a^2\mathcal{J}S^2}{4}\sum_{i}\left(\vb{\grad}\vb{n}(\vb{r}_i)\right)^2.
\label{HamJ3}
\end{split}
\end{equation}
To obtain the continuum limit, we make the substitution
\begin{equation}
\begin{split}
\sum_if_i\rightarrow \frac{1}{\Omega}\int_\Omega\dd^dr\;f(\vb{r}),
\label{Substi}
\end{split}
\end{equation}
eventually leading to 
\begin{equation}
\begin{split}
H=\int_{A}\dd^2r\;\left[\frac{3a^2\mathcal{J}S^2}{4A}\left(\vb{\grad}\vb{n}(\vb{r})\right)^2-\frac{2\mathcal{K}S^2}{2A}n_z^2(\vb{r})\right]
\label{HamConti1}
\end{split}
\end{equation}
for the Hamiltonian, where $A$ is the unit cell area of the plane perpendicular to the anisotropy direction. 
Equating the above with the continuum limit Hamiltonian in Eq. (4) of the main text, we obtain 
the relation between the exchange coupling and the on-site magnetic anisotropy in the two models
\begin{align}
	&J=\frac{3a^2\mathcal{J}S^2}{2A},\\
	&K=\frac{2\mathcal{K}S^2}{A}.
	\label{Constants}
\end{align}
\section{Material specific parameters and computational details}
\label{SuppInfo-MatSpec-SubSec}
\subsubsection{{MnBi$_2$Te$_4$}}
Density functional theory (DFT) calculations were performed based on experimental bulk crystal structure
of {\mbt} using the GGA+$U$ method with the Perdew-Burke-Ernzerhof
(PBE) approach 
\cite{perdew1996generalized} as implemented in the FPLO code version 18.00-52
\cite{PhysRevB.59.1743,fplo_web}. We used the atomic limit double counting correction and tetrahedron method for $k$-space integrations. For the slab calculation, we used a mesh of $12\times12\times1$ subdivisions in the Brillouin zone (and $36\times36\times1$ for the density of states calculation), while for the bulk calculations we used $16\times16\times16$ (rhombohedral setup).

To estimate the magnetic anisotropy $K$, we did calculations for the bulk system, ferromagnetic
configuration, with quantization axis along [001] or [100]. The results
are sensitive to the values of $U$ and $J$ used to treat electronic correlations in the
Mn-$3d$ shell.
Varying $U-J$ between 1 and 5.34\,eV, we find that the magnetic anisotropy energy varies between 0.46\,meV
and 0.27\,meV per Mn or, accordingly, the on-site anisotropy for the Heisenberg model with $S=5/2$,
$\mathcal{K}$,  between $0.073$\,meV and $0.043$\,meV.
Similarly, based on additional calculations where the Mn are ordered ferromagnetically between layers and antiferromagnetically within layers, we estimate the intralayer exchange coupling $\mathcal{J}$. 
In the same range of $U-J$ as above, we find $\mathcal{J}$ to vary between 0.18\,meV and 0.5\,meV.
The trends are similar to those reported in Ref. \cite{Li-McQueeney-PRL-AFMTIMnBi2Te4-ExchangeandAniso-2020}.
For the Monte Carlo simulations, we used the values obtained with
$U-J=1$, which yield the ratio $\mathcal{K}/\mathcal{J} \sim 0.4$, in very good agreement with Ref. \cite{Li-McQueeney-PRL-AFMTIMnBi2Te4-ExchangeandAniso-2020}.

\begin{figure}[h!]
	\centering
	\includegraphics[width=2.8 cm]{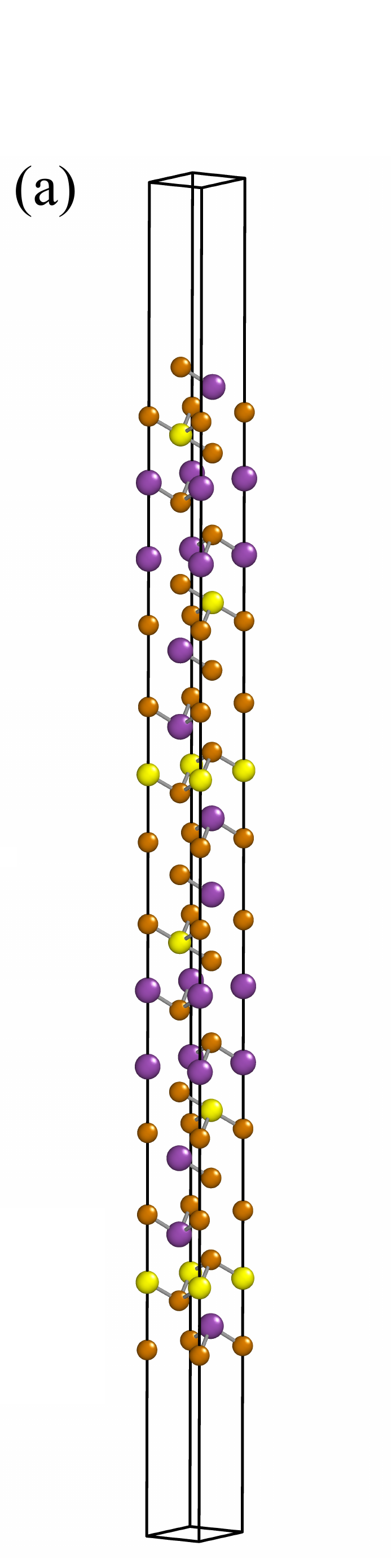}
	\includegraphics[width=9 cm]{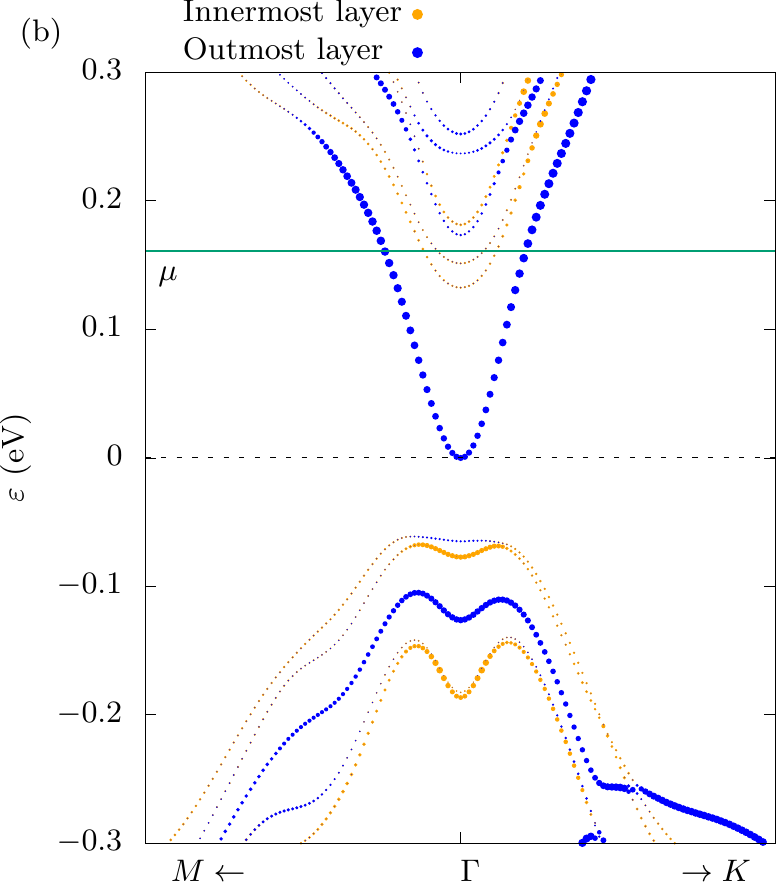}
	\caption{(a) Slab used for the DFT calculation. (b) Layer-projected band structure. $\mu$
		indicates the estimated chemical potential, based on the density of states of the slab and the
		carrier density reported in Ref. \cite{Chulkov-Isaeva}. Yellow atoms correspond to Mn, while orange to Bi and violet to Te. }
	\label{fig:dft_bands_zoom}
\end{figure}

To estimate the Fermi velocity of the surface state and the surface gap, we performed  DFT
calculations for a slab structure consisting of six MnBi$_2$Te$_4$ unit cells with a vacuum of 30
Bohr radii [Fig. \ref{fig:dft_bands_zoom}(a)]. 
Following Ref. \cite{Chulkov-Isaeva,vidal2020orbital}, we fix $U=5.34\,$eV and $J=0$.
The two main surface bands present a gap $\sim$100\,meV, as shown in Fig.
\ref{fig:dft_bands_zoom}(b), in good agreement with the value  of
88\,meV found in \cite{Chulkov-Isaeva}. The Fermi velocity found for the upper part of the
Dirac cone is approximately $2.3\pm0.3$\,eV{\AA} in good agreement with the experimental results in Ref.
\cite{PhysRevX.9.041040} and DFT results in Ref. \cite{Chulkov-Isaeva}.

Lastly, samples of MnBi$_2$Te$_4$ tend to be self-doped, meaning that different kind of defects place the chemical potential $\mu$ outside the gap. In particular, the samples tend to be electron-doped.
To estimate the value of $\mu$, we consider the estimation of carriers of $n_{\rm c}=2\times10^{19}/$cm$^3$ provided in Ref. \cite{Chulkov-Isaeva}. Based on this value and on our slab calculation we estimate $\mu\sim160\,$meV above the bottom of the conduction band.

Notice that the value of $\mu=160$ meV with respect to the bottom of the conduction band corresponds to $210$ meV with respect to our zero of energies, as we have to add half the size of the gap.

\subsubsection{{EuS-Bi$_2$Se$_3$} heterostructures}

In {\eus} heterostructures, the interface is typically formed by the $(111)$ surface of the cubic
bulk-EuS structure such that the lattice mismatch with the topological Bi$_2$Se$_3$ film is
minimal \cite{PrezVicente-BiSeLatticeConst-1999}. 
For this surface, the interfacial layer of Eu atoms span a triangular lattice. The effective 
lattice constant of this lattice is 
$a=a_{\text{EuS}}/\sqrt{2}\approx 4.22\;\angstrom$, using $a_{\text{EuS}}\approx 5.96\;\angstrom $
\cite{Wachter-EuSLatticeConst-1972}.

The Fermi velocity was obtained from \cite{Zhang-Cheng-nature-BiSeFermiVelocity-2009} to approximately be $\hbar v_{\textrm{F}}\approx 3.29\;\textrm{eV \angstrom}$ for bulk $\text{Bi}_2\text{Se}_3$. Furthermore, the exchange coupling constant $J_0$ was estimated in alignment with the magnetic gap reported in \cite{Wei-Katmis-Moodera-PRL-2013,Eremeev-Chulkov-EuSBiSe-2015} to have a value of $54$ meV.

Regarding magnetism in the {\eus}, the value of
$\mathcal{J} = 0.017\,$meV has been reported earlier \cite{Mauger-Godart-EuS-1986}. 
The value of the on-site magnetic anisotropy was obtained from the EuS layer thickness
dependence of the magnetic anisotropy \cite{Nogueira-Katmis-Moodera-Nature-2016-TcShift}. We
considered the structure with the largest Bi$_2$Se$_3$ layer thickness of 20 nm. Note that
these values are available in the continuum limit and were converted to the lattice equivalent
values using Eq. (\ref{Constants}). The resulting data was modeled with \cite{Story-EuSPbSAnistropyThicknessScaling-2000}:
\beq
K_1(d) = K_{\textrm{V}} + \frac{2K_{\textrm{S}}}{d}\,,
\eeq
where $K_{\textrm{V}}$ and $K_{\textrm{S}}$, respectively denote the bulk and surface magnetic
anisotropy contributions and $d$ is the thickness of the EuS layer. 
For the EuS monolayer along the $(111)$-direction, with a thickness of $d=a_{\textrm{EuS}}/\sqrt{3}\approx 3.45\;\angstrom$, we obtain $\mathcal{K}
\approx 0.126\,$meV, 
leading to $\mathcal{K}/\mathcal{J} \approx 7.4$.

\subsubsection{Monte-Carlo calculations of {$T_{\rm c}$}}

Classical Monte-Carlo simulations (MCS) with the Metropolis algorithm were carried out for spins on a two-dimensional triangular
lattice with $42 \times 42$ sites. 
We consider the spin Hamiltonian of Eq. (\ref{AppHamHeisenberg}).
$\mathcal{J}S^2 = 1$ defines the energy scale leaving the ratio $\mathcal{K}/\mathcal{J}$ as the
only free parameter.
For each $\mathcal{K}/\mathcal{J}$, we started from a high-temperature paramagnetic state, characterized by a random spin configuration,
and decreased the temperature in steps of 0.02. At each temperature, 
the system was allowed to equilibrate over $N_{\rm eq}$ steps and the physical quantities
were obtained by averaging over the next $N_{\rm av}$ steps. For $\mathcal{K}/\mathcal{J} < 1$, the
equilibration was typically reached in $\lesssim 5 \times 10^4$, however, to treat the entire range of
$\mathcal{K}/\mathcal{J} < 10$ on same footing, we generously consider $N_{\rm eq} = 2 \times 10^{5}$ and 
$N_{\rm av} = 3 \times 10^{5}$ steps.
The critical temperature, $T_{\rm
	c}^{\rm latt}$, was obtained from peak(s) in the specific heat, which agrees with corresponding values obtained from the magnetization data $M$ vs. $T$.

\begin{figure}[ht!]
	\centering
	\includegraphics[width=0.48\columnwidth]{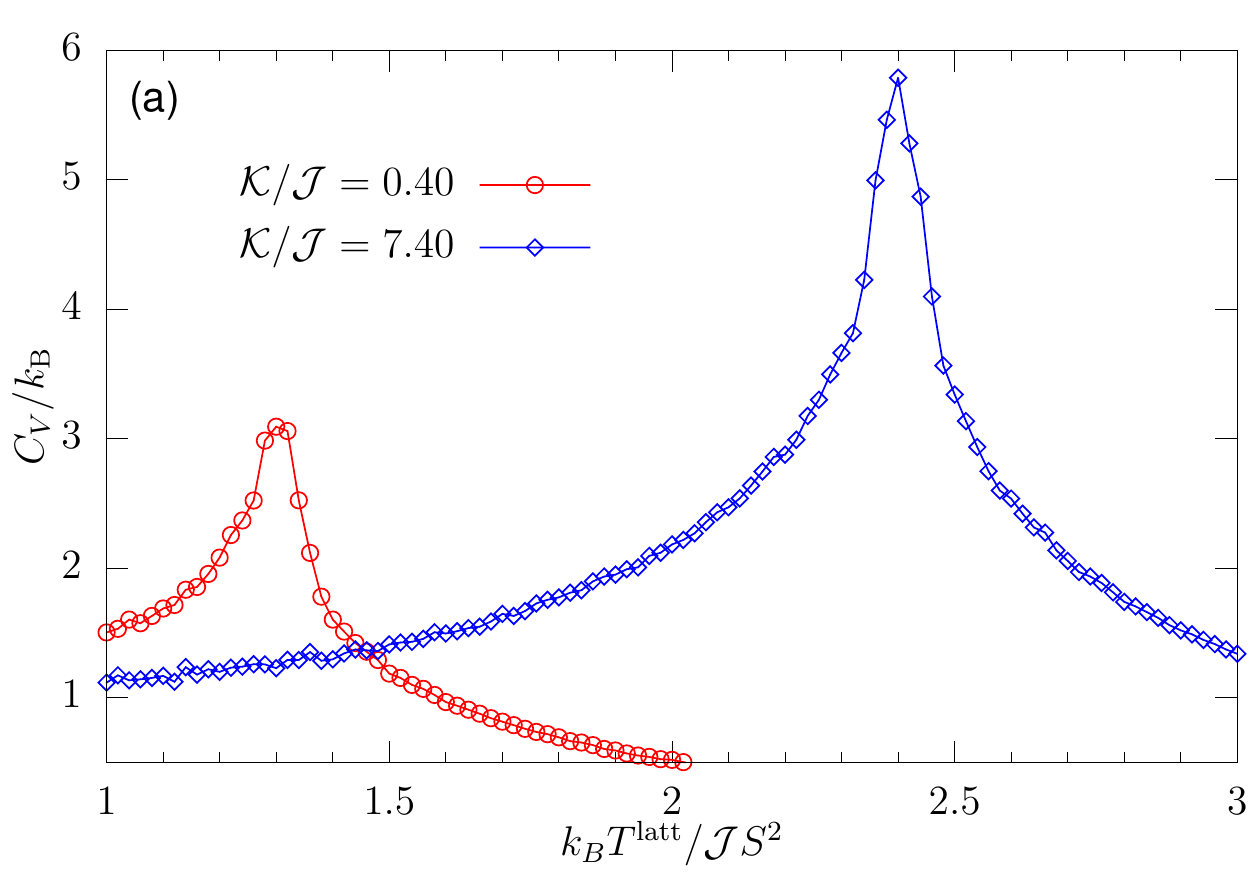}
	\includegraphics[width=0.48\columnwidth]{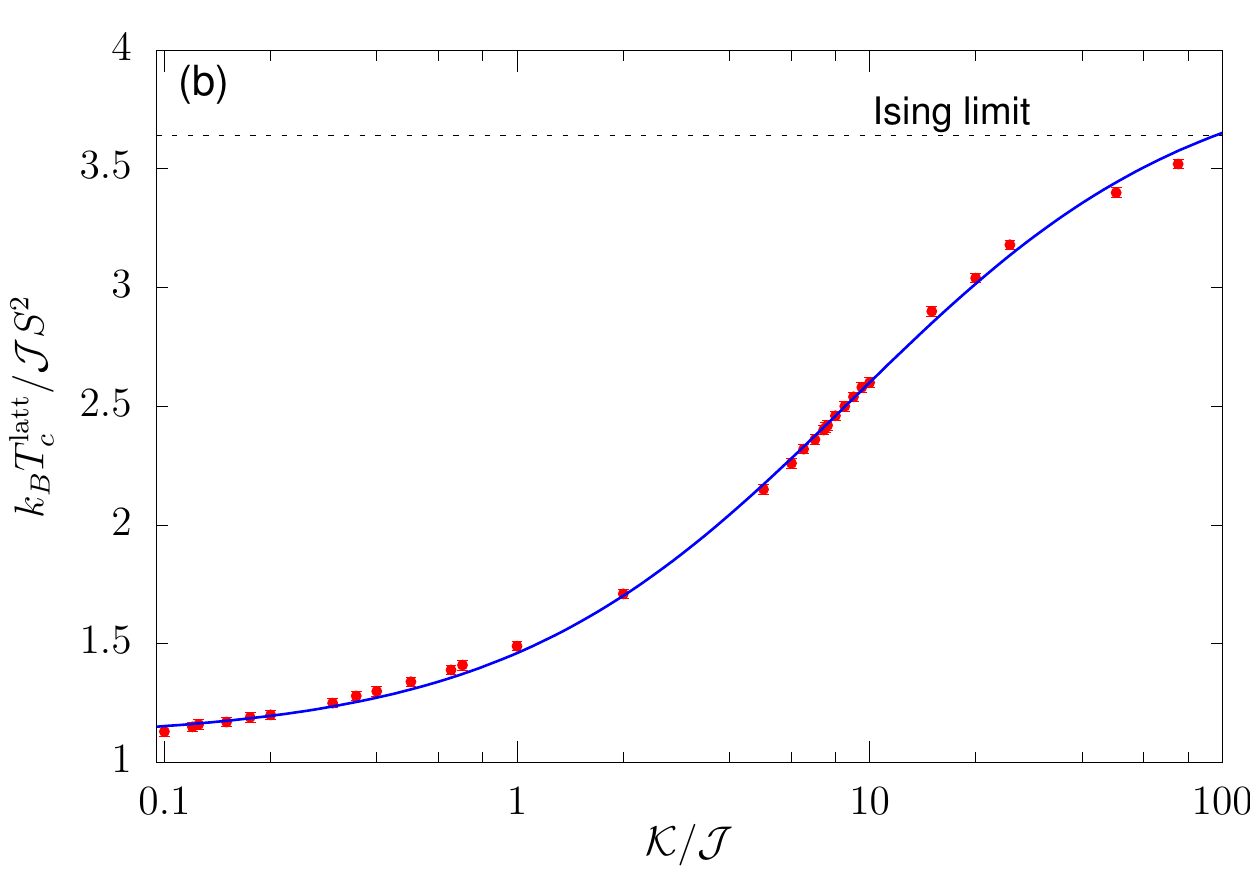}
	\caption{(a) Specific heat for the values of $\mathcal{K}/\mathcal{J}$ corresponding to the
		monolayer of {\mbt} and {\eus} heterostructure. (b) $T_{\rm c}^{\rm latt}$ for a wide range of
		parameter, showing the evolution to the Ising limit (dashed line) for $\mathcal{K} \gg
		\mathcal{J}$. The solid line is a guide to the eye.
	}
	\label{fig:mcs}
\end{figure}

To  
address the materials of interest in this study, we used
$\mathcal{K}/\mathcal{J} = 0.40$ for {\mbt} and $\mathcal{K}/\mathcal{J} = 7.4$
for the {\eus} heterostructure as discussed earlier. 
Figure \ref{fig:mcs} (a) shows the corresponding specific heat data. 
From the well-defined peaks, we obtain $k_{\rm B}T_{\rm c}^{\rm latt}/\mathcal{J}S^2 \sim 1.46\,$ and
$\sim 2.40$, for the {\mbt} and {\eus} monolayers, respectively. Considering $S=5/2$ for {\mbt}, we
obtain $T_{\rm c}^{\rm latt} \sim 16.97\,$K, while for the {\eus} heterostructure, $S=7/2$ yields
$\sim 5.80\,$K.

For completeness, 
we also carried out MCS over a wide range of $\mathcal{K}/\mathcal{J}$.  
This allows us to study the evolution of $T_{\rm c}^{\rm latt}$ as a function of $\mathcal{K}/\mathcal{J}$
and to analyze how close or far are the systems of interest from the Ising limit $\mathcal{K}/\mathcal{J} \gg 1$. 
With increasing $\mathcal{K}/\mathcal{J}$, the accessible phase space becomes considerably smaller. Consequently, the equilibration is slower.  
Therefore, for $\mathcal{K}/\mathcal{J} \geq 10$, MCS were
carried out with a total of $1 \times 10^6$ update steps, out
of which the first $5 \times 10^5$ steps correspond to $N_{\rm eq}$ and were discarded during the averaging and evaluation of the
physical properties such as specific heat.
Figure \ref{fig:mcs} (b) shows the evolution of $T_{\rm c}^{\rm latt}$ with $\mathcal{K}/\mathcal{J}$. The Ising limit, which 
corresponds to $k_{\rm B}T_{\rm c}^{\rm latt}/\mathcal{J}S^2 = 3.642$ \cite{ghaemi2001}, is approximately 
reached for $\mathcal{K}/\mathcal{J} \gtrsim 100$.

\subsection{Fluctuations away from the saddle point}
\label{SuppInfo-Fluctuations-SubSec}
Starting with the same model Hamiltonian but now also accounting for fluctuations around the previously analyzed saddle-point we assume the magnetization to have the form $\vb{n}(\vb{r},t)=n_z\vu{z}+\delta\vb{n}(\vb{r},t)$ and also for there to be an electric potential $\phi$, which is of either external or internal origin.
The dimensionless euclidean action of our partition function is then given by
\begin{equation}
\begin{split}
\hat{S}_{\textrm{F}}&=\frac{1}{\hbar}\int_0^{\hbar\beta}\dd \tau\int\dd^2r\;{\Psi}^\dagger\big[\hbar\partial_\tau+H_{\rm{Dirac}}\big]\Psi\\
&=\frac{1}{\hbar^3\beta^2}\sum_{n,m}\int\frac{\dd^2q}{(2\pi)^2}\int\frac{\dd^2k}{(2\pi)^2}\braket{\vb{k},n}{\vb{q},m}\Psi^\dagger_{\vb{k},n}\left[-i\hbar\omega_m-\mu+\vb{d}(\vb{q})\cdot\pmb{\sigma}-m\sigma_z\right]\Psi_{\vb{q},m}\\
&\quad+\frac{1}{\hbar^3\beta^2}\sum_{n,m}\int\frac{\dd^2q}{(2\pi)^2}\int\frac{\dd^2k}{(2\pi)^2}\Psi^\dagger_{\vb{k},n}\left[-e\phi(\vb{k}-\vb{q},i\nu_{n-m})-J_0\delta\vb{n}(\vb{k}-\vb{q},i\nu_{n-m})\cdot\pmb{\sigma}\right]\Psi_{\vb{q},m}\\
&=\frac{1}{\hbar^2\beta^2}\sum_{n,m}\int\frac{\dd^2q}{(2\pi)^2}\int\frac{\dd^2k}{(2\pi)^2}\Psi^\dagger_{\vb{k},n}\mel{\vb{k},n}{-\mathbb{G}^{-1}+\mathbb{V}}{\vb{q},m}\Psi_{\vb{q},m}\\
&=\frac{1}{\hbar^2\beta^2}\sum_{n,m}\int\frac{\dd^2q}{(2\pi)^2}\int\frac{\dd^2k}{(2\pi)^2}\Psi^\dagger_{\vb{k},n}\mel{\vb{k},n}{-\mathbb{G}^{-1}\big(\mathbbm{1}-\mathbb{G}\mathbb{V}\big)}{\vb{q},m}\Psi_{\vb{q},m},
\label{App:Eq:PartDef1}
\end{split}
\end{equation}
where we defined the bosonic Matsubara frequencies  $\nu_n=2n\pi/\hbar\beta$ and
\begin{align}
	&\mel{\vb{k},n}{-\mathbb{G}^{-1}}{\vb{q},m}=\frac{1}{\hbar}\braket{\vb{k},n}{\vb{q},m}\left[-i\omega_m-\mu+\hbar v_{\textrm{F}}\vb{d}(\vb{q})\cdot\pmb{\sigma}-m\sigma_z\right],\label{App:Eq:PartCalc2.1}\\
	&\mel{\vb{k},n}{\mathbb{V}}{\vb{q},m}=-\frac{1}{ \hbar}\left[e\phi(\vb{k}-\vb{q},i\nu_{n-m})+J_0\delta\vb{n}(\vb{k}-\vb{q},i\nu_{n-m})\cdot\pmb{\sigma}\right]=:-\frac{1}{\hbar}\hat{\mathcal{V}}(\vb{k}-\vb{q},i\nu_{n-m}).
	\label{App:Eq:PartCalc2.2}
\end{align}
Performing the path integration over the fermionic fields in the saddle-point approximation yields the fermionic effective action up to second order
\begin{equation}
\begin{split}
\hat{S}_{\rm{F},\rm{eff}}&=\frac{1}{2}\Tr\Big[\mathbb{G}\mathbb{V}\mathbb{G}\mathbb{V}\Big]\\
&=\frac{1}{2\hbar^2\beta^2}\sum_{n,m}\int\frac{\dd^2q}{(2\pi)^2}\int\frac{\dd^2k}{(2\pi)^2}\Big[\mathcal{G}_{\vb{k},n}^{\alpha\beta}\mathcal{G}_{\vb{q}+\vb{k},m+n}^{\gamma\delta}\mathcal{V}_{-\vb{q},-m}^{\beta\gamma}\mathcal{V}_{\vb{q},m}^{\delta\alpha}\Big],
\end{split}
\label{App:Eq:ActCalc1}
\end{equation}
where the Greek indices label the spin components resulting from the trace over the Pauli matrices and where we defined
\begin{equation}
\begin{split}
&\hat{\mathcal{G}}_{\vb{k},n}=\frac{(i\hbar\omega_n+\mu)+\hbar v_{\textrm{F}}\vb{d}(\vb{k})\cdot\pmb{\sigma}-m\sigma_z}{(i\hbar\omega_n+\mu)^2-\hbar^2 v^2_{\textrm{F}}\vb{d}^2(\vb{k})-m^2}.\\
\end{split}
\label{App:Eq:GFDef2}
\end{equation}
By insertion of the definition of $\mathcal{V}$ into Eq. (\ref{App:Eq:ActCalc1}) one can now split up the different parts of the action according to their magnetoelectric nature, yielding an electric, magnetic and magnetoelectric contribution.

To arrive at the DMI and CS terms we then trace out the spin components, and calculate the integral over $\vb{k}$ and the sum over fermionic Matsubara frequencies $i\omega_n$ in a derivative expansion, that means a long wavelength and low frequency expansion in terms of the wavevector $\vb{q}$ and bosonic Matsubara frequencies $i\nu_m$. This step is straightforward but very lengthy and shall be omitted here.

Afterwards one transforms back to Euclidean spacetime using the remaining Matsubara sum and momentum integral over $i\nu_m$ and $\vb{q}$ to find the result for the DMI and CS terms presented in the main text, among other contributions.

\end{document}